\documentclass{pasj01}
\Received{$\langle$reception date$\rangle$}
\Accepted{$\langle$acception date$\rangle$}
\Published{$\langle$publication date$\rangle$}

\usepackage{ulem}
\usepackage{color}

\begin{document}

\title{Hoyle-Lyttleton accretion on to black-hole accretion disks with super-Eddington luminosity for dusty-gas}
\author{Erika \textsc{Ogata},$^{*}$
        Ken \textsc{Ohsuga},
        Hidenobu \textsc{Yajima}$^{\dag}$ }%
\altaffiltext{$\dag$}{Center for Computational Sciences, 
                University of Tsukuba, 
                Ten-nodai, 1-1-1 Tsukuba, 
                Ibaraki 305-8577, Japan}
\email{ogata@ccs.tsukuba.ac.jp}


\maketitle

\begin{abstract}
  We investigate the Hoyle-Lyttleton accretion of dusty-gas for the case 
  where the central source is the black hole  accretion disk. 
  By solving the equation of motion taking into account the radiation force 
  which is attenuated by the dust absorption, 
  we reveal the steady structure of the flow around the central object. 
  We find that the mass accretion rate tends to increase with an increase of the optical thickness of the flow 
  and the gas can accrete even if the disk luminosity exceeds the Eddington luminosity for the dusty-gas, 
  since the radiation force is weakened by the attenuation via the dust absorption. 
  When the gas flows in from the direction of the rotation axis for the disk with $\Gamma'=3.0$, 
  the accretion rate is about 93\% of the Hoyle-Lyttleton accretion rate 
  if $\tau_{\rm HL}=3.3$ and zero for $\tau_{\rm HL}=1.0$, 
  where $\Gamma'$ is the Eddington ratio for the dusty-gas 
  and $\tau_{\rm HL}$ is the typical optical thickness of the Hoyle-Lyttleton radius.
  Since the radiation flux 
  in the direction of disk plane is small, 
  the radiation force tends not to prevent gas accretion from the direction near the disk plane. 
  For $\tau_{\rm HL}=3.3$ and $\Gamma'=3.4$, 
  although the accretion is impossible in the case of $\Theta=0^{\circ}$, 
  the accretion rate is 28\% of the Hoyle-Lyttleton one in the case of $\Theta=90^{\circ}$,
  where $\Theta$ is the angle between the direction the gas is coming from and the rotation axis of the disk.
  We also obtain relatively high accretion luminosity that is realized when the accretion rate of the disk onto the BH is consistent with that via the Hoyle-Lyttleton mechanism taking into account the effect of radiation.
  This implies the intermediate-mass black holes moving 
  in the dense dusty-gas are identified as luminous objects in the infrared band.

\end{abstract}

\section{Introduction}
\label{Introduction}

Revealing the formation and growth of black holes (BHs) is one of the central issues in astronomy.
Recent observations have successfully detected several tens of stellar-mass BHs in our galaxy as X-ray binaries
\citep{corral2016}.
However, considering the total stellar mass and the initial mass function, more BHs could have been formed in our galaxy up to the present day.  
\citet{agol2002} suggested that our galaxy might harbor $\sim 10^{8}-10^{9}$ BHs totally (see also, \cite{caputo2017}). 
Recent observations have indicated
isolated BHs might be floating in interstellar space 
(\cite{sashida2013}; \cite{oka2016}; \cite{oka2017}; \cite{takekawa2017}; \cite{yamada2017}; \cite{takekawa2019}).
The isolated BHs can
attract and swallow the gas 
while moving in interstellar space.
This process is known as the Hoyle-Lyttleton accretion (\cite{hoyle1939}), 
and plays a main roll in the growth of BHs.
In fact, 
\citet{rice2020}, \citet{safarzadeh2020} 
proposed that the gas accretion via 
the Bondi-Hoyle-Lyttloton mechanism may solve the mass gap problem
pointed out by
\citet{abbott2020}, \citet{liu2020} 
from 
the detection of the gravitational wave.
However, the spatial distribution of BHs and their growth mechanism have not been understood yet.

The study of the gas accretion onto a BH 
was initiated late 1930s
(\cite{hoyle1939}; \cite{bondi1944}; \cite{bondi1952}).
Subsequently, a detailed investigation 
was conducted using the hydrodynamics simulations 
(\cite{shima1985}; \cite{ruffert1994}; \cite{ruffert1996}).
These previous studies did not take into account the radiation force that changes the gas dynamics significantly as the source becomes bright. 
Therefore, the radiation hydrodynamic simulations of the gas accretion were performed with the assumption of an isotropic central source recently
(\cite{milosavljevic2009}; Park \& Ricotti \yearcite{park2011},\yearcite{park2012},\yearcite{park2013}; \cite{sugimura2020}; \cite{toyouchi2020}).
When the accretion disk forms around the BH, 
the radiation field is expected to be anisotropic.
It allows the gas accretion from a shadow region, resulting in a higher accretion rate  (e.g., \cite{sugimura2017}; \cite{takeo2018}).

The radiation hydrodynamics simulations are still expensive to investigate various parameters as gas density, flow speed, and the anisotropy of radiation. Therefore the dependencies on the parameters are poorly understood.
An alternative way is to solve the steady flow structure of the accreting gas, of which the low computational cost allows the systematic study.
Based on the analysis of the steady flow, 
\citet{fukue1999} examined the Hoyle-Lyttleton accretion 
around BH accretion disks with the luminosities lower than the Eddington limit.
They revealed that 
the accretion rate became small 
as the Eddington ratio increased,  and the reduction rate could be more than 70\%, compared to the cases without the radiation feedback.
\citet{hanamoto2001} calculated the Hoyle-Lyttleton accretion onto the super-Eddington disks.
They showed that 
the self-shielding of the disk drastically reduces the radiation force,
resulting in the high accretion rate even for the super-Eddington luminosity. 

The above studies of the steady flow ignored interstellar dust. As star formation proceeds, galaxies become enriched with dust and metals. The dust-to-gas mass ratio is $\sim 1\%$ in our galaxy (\cite{zubko2004}) and higher in quasars or ultraluminous infrared galaxies (\cite{solomon1997}). In such a situation, the radiation force on the dust has significant impacts on the gas accretion because of the higher absorption efficiency of dust (e.g., \cite{yajima2017}). 
In the case without the dust, the radiation force on electrons simply decreases with the square of the distance from a source as the gravitational force in the rarefied plasma. On the other hand, the radiation force on dust also changes due to the dust attenuation, which makes the analysis of the flow structure complicated.

Thus, in this study, we investigate the steady structure 
of Hoyle-Lyttleton accretion of dusty-gas with the radiative transfer considering the dust attenuation. 
For this purpose, 
the equation of motion, 
which incorporate the gravity of the BH and the radiation force by the BH accretion disk, 
is solved combined with the continuity equation.
Here, we assume that the gas pressure is negligibly small since the velocity is larger than the sound velocity. However, this situation may change if the shock occurs. We will discuss shock wave in detail \S \ref{Future work}.
As a result, 
we can evaluate the accretion rate in the situation where BHs are moving in the high-density dusty-gas.
This paper is organized as follows.
In \S \ref{Basic eq}, we introduce basic equations and describe the numerical method.
In \S \ref{Result}, we show our simulation results and \S \ref{Discussion} is devoted to discussion.
Finally, the summary and conclusion are given in \S \ref{Conclusion}.

\section{Basic Equations and Numerical Method}
\label{Basic eq}
\subsection{overview}
\label{overview}
We study a steady structure of flows around the BHs
which are surrounded by the accretion disks
and moves in the dusty-gas.
The method for investigating the structure of the flow
is similar to that used by \citet{fukue1999}
but we consider the extinction of the radiation
from the accretion disks via the absorption by the dusty-gas.
The steady structure of the flows is given by the calculation
of the trajectories of fluid elements (streamlines),
which come towards the BHs from far enough away
in the BH rest frame.
We solve the equation of motion
considering the radiation force as well as the gravity
with the mass conservation along the streamline.

\begin{figure*}[t]
  \begin{center}
    \includegraphics[width=150mm]{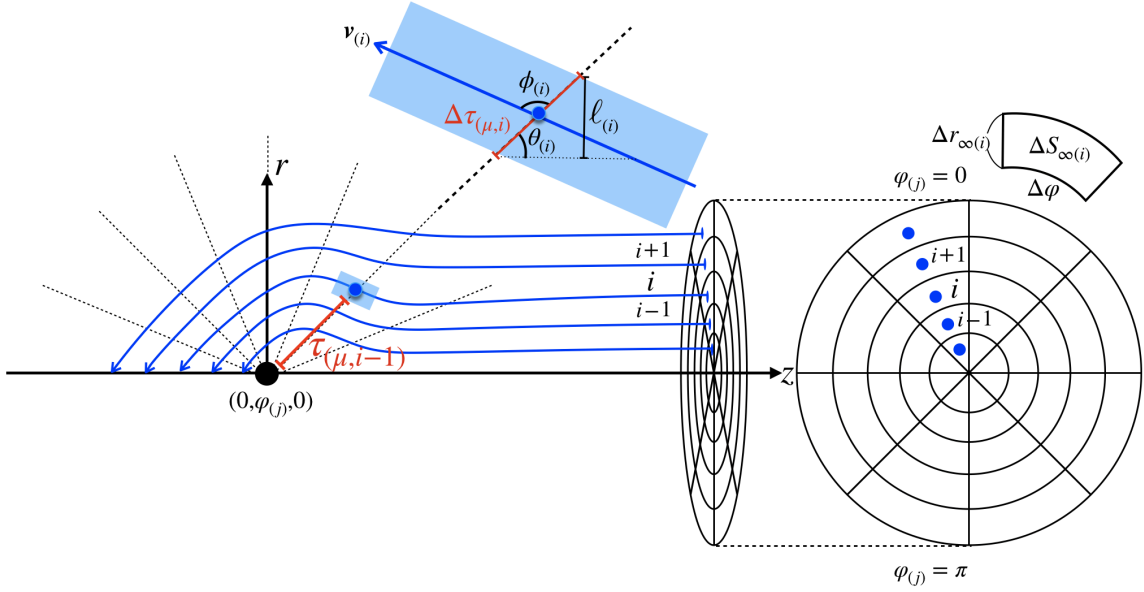}
  \end{center}
  \caption{
  Configuration of the system.
  We adopt cylindrical coordinates $(r,\varphi,z)$ and the BH accretion disk is located at the origin. The blue arrows indicate streamlines in the plane where phi is constant, and the blue point in the right circle is the starting points of the calculation of the streamlines.  The optical depth is measured along radial dashed lines extending from the origin. The radiation force at the blue point on the $i$-th streamline is calculated using the optical depth due to the streamlines up to the ($i-1$)-th, $\tau_{(m,i-1)}$, and the optical depth of the $i$-th streamline itself, $\Delta\tau_{(m,i)}$. The top center figure shows an enlarged view of the blue point on the $i$-th streamline. Here, $\boldsymbol{v}_{(i)}$ is the velocity vector of streamline, $\phi_{(i)}$ is the angle between velocity vector and position vector, $\theta_{(i)}$ is the polar angle measured from $z$-axis, and $l_{(i)}/\rm{sin}\theta_{(i)}$ is the path length as photon passes through the $i$-th streamline. 
  
  }\label{fig1}
\end{figure*}

Figure \ref{fig1} shows configuration of the system.
We use the cylindrical coordinates $(r,\varphi,z)$
where the central BH is located at the origin.
The accretion disk around the BH is small enough to be recognized
as the point-like radiation source.
The rotation axis of the disk is located
on the plane of $\varphi=0$,
and the angle between the rotation axis
and the $z$-axis is $\Theta$.
All streamlines are supposed to be parallel to the $z$-axis
at the starting point of the streamlines, $z=z_{\rm ini} \gg R_{\rm HL}$
where $R_{\rm HL}$ is the Hoyle-Lyttleton accretion radius.
At the starting point, the density and the velocity of the flow are
assumed to be uniform.
When we solve $i$-th streamline,
the radiation from the central object
is diluted via the absorption by the $i'$-th ($1\leq i'<i$) streamlines.
We take into account of the self-shielding by $i$-th flow.
When the calculation of the final (outermost) streamline is finished,
the global structure of the flow is obtained.

At the point where the streamline reaches $z$-axis behind the BH ($z<0$),
we check whether the condition for the accretion is satisfied or not.
Using the streamlines which meet the condition,
we evaluate the mass accretion rate onto the center, $\dot{M}$.
In \S\ref{Numerical Method}, numerical method is described in more detailed.

\subsection{Basic Equations}
As we mentioned above,
we use the cylindrical coordinates
and solve the equation of motion for the dusty-gas,
\begin{eqnarray}
  \frac{d\boldsymbol{{v}}}{d{t}}&=&-\frac{GM\boldsymbol{{R}}}{{R}^3}+\boldsymbol{f}_{\rm{rad}},
  \label{eq.motion}
\end{eqnarray}
where $G$ is the gravitational constant, $t$ is the time,
$M$ is the mass of the BH,
$\boldsymbol{v}=(v^r, v^\varphi, v^z)$
and $\boldsymbol{{R}}$ are the
velocity and the position of the fluid element,
$R=\sqrt{r^2+z^2}$ is the distance from the origin,
and $\boldsymbol{f}_{\rm{rad}}$ is the radiation force
acting on the dusty-gas.
We assume that the gas pressure is negligibly small
(we will discuss later).

In the present study,
the opacity of the dusty-gas is given by
\begin{eqnarray}
  \kappa_{\rm{dg}}&=&
  \kappa_{\rm{es}}+\alpha f_{\rm{dg}}\kappa_{\rm{dust}},
  \label{opaci}
\end{eqnarray}
where $\kappa_{\rm{es}}\sim 0.4\;\rm{cm^2\;g^{-1}}$ is the electron scattering opacity,
$f_{\rm{dg}}$ is the dust-to-gas mass ratio,
$\kappa_{\rm{dust}}$ is the absorption opacity of the dust
estimated by geometrical cross section,
and $\alpha$ is constant less than unity,
determined by the absorption coefficient to geometrical cross section
and the spectral energy distribution of the accretion disk.
In equation (\ref{opaci}), the dusty-gas is supposed to be ionized. Although this is useful in the vicinity of the radiation source, it is thought to overestimate in the neutral region where the radiation is sufficiently attenuated due to dust absorption. However, the opacity of the electron scattering is sufficiently smaller than the opacity of the dust that this overestimation is negligibly small. If we employ the typical values,
$\kappa_{\rm{dust}} \sim 0.75\times10^5\; \rm{cm^2\;g^{-1}}$,
$f_{\rm{dg}} \sim 0.01$,
and $\alpha \sim 0.1$,
we have $\kappa_{\rm es}/\kappa_{\rm dust}\sim 5.3\times10^{-3}$. Thus, we employ equation (\ref{opaci}) regardless of the ionization structure throughout the present study.
Here $\alpha$ and $f_{\rm dg}$ dependencies
will be discussed later.
In addition, 
we assume that dust is not sublimated 
throughout the present study.
This validity will be discussed in \S \ref{Discussion}.
Although the dust would emit the infrared radiation,
the opacity of the dust is thought to be very small
at the infrared band.
Thus, we neglect the reprocessed radiation
(scattered radiation and re-emitted radiation) in the present study.

Since the accretion disk around the BH is recognized as the point source,
and since the reprocessed radiation is neglected,
only the $R$-component of the radiation flux ($F_{\rm rad}$) is not zero.
If the accretion disk is optically thick and geometrically thin,
the radiation flux is evaluated as $F_{\rm rad}=L  \cos\psi e^{-\tau}/2\pi R^2$
where
$\tau$ is the optical depth measured from the origin,
$L$ is the luminosity of the accretion disks,
$\psi$ is the angle from the rotation axis of the disk,
$\cos\psi=|\boldsymbol{n}_{\rm rot} \cdot \boldsymbol{R}|/R$,
with $\boldsymbol{n}_{\rm rot}$ being the unit vector in
the direction of the rotation axis.
For comparison, we also consider the case of the isotropic radiation
(isotropic model), $F_{\rm rad}=L e^{-\tau}/4\pi R^2$.

The $R$-component of the radiation force is
\begin{eqnarray}
  f^R_{\rm{rad}}=\frac{\kappa_{\rm dg}}{c} F_{\rm rad} \frac{1-e^{-\Delta\tau}}{\Delta\tau},
\end{eqnarray}
where $c$ is the speed of light,
and $\Delta\tau$ is the optical depth of the streamline
measured in the $R$-direction.
The optical depth is evaluated with using the
mass conservation law along the streamlines
(we will explain the detailed evaluation method
in \S\ref{Numerical Method}).
As with the radiation flux,
only the $R$-component of the radiation force is non-zero.

Using the effective Eddington ratio for the dusty-gas,
$\Gamma'=\kappa_{\rm dg} L/\kappa_{\rm es} L_{\rm Edd}$
with $L_{\rm Edd}$ being the Eddington luminosity,
the equation (\ref{eq.motion}) is rewritten as
\begin{eqnarray}
  \frac{d\boldsymbol{{v}}}{d{t}}
  =&-&\frac{GM \boldsymbol{{R}}}{{R}^3}\nonumber\\ 
  &\times&
  \left\{ 1-\Gamma' e^{-\tau}\frac{1-e^{-\Delta\tau}}{\Delta\tau}
  \left[ \left(2\cos\psi -1\right) \delta+1 \right] \right\},
  \label{eq.motion2}
\end{eqnarray}
where $\delta$ is null for the isotropic model
and unity for the disk case.

\subsection{Numerical Method}
\label{Numerical Method}
The right figure in figure \ref{fig1} shows the cross-sectional view at $z=z_{\rm ini}$.
We calculate trajectories of the fluid element (streamlines)
passing through the half of the circle ($\varphi=0-\pi$) with the radius $R_{\rm HL}$ 
since the flow is symmetric with respect to
the $\varphi=0$ plane (or $\varphi=\pi$ plane).
At $z=z_{\rm ini}$,
we suppose the dusty-gas of uniform density flows in the $z$-direction.
Hereafter the quantities at $z=z_{\rm ini}$ are denoted
by the subscript $\infty$ as $\rho_\infty$ and $v^z_\infty$.
We divide the half circle into $540 \times 40$ grid cells.
The grid spacing in the $r$- and $\varphi$-directions
is set to be constant,
$\Delta r=R_{\rm HL}/540$ and $\Delta \varphi=\pi/40$, respectively.
We calculate $540 \times 40$ trajectories of the fluid element (streamlines)
by solving the equation (\ref{eq.motion2})
with using the fourth-order Runge-Kutta method.

Since $\varphi$-component is zero in both radiation force and gravity,
and since the $r$- and $\varphi$-components of the velocity is zero at $z=z_{\rm ini}$,
the fluid element moves on the plane of $\varphi$ is constant.
Thus, streamlines in the plane of $\varphi=\varphi_{(j)}$
and those in the plane of $\varphi=\varphi_{(j'\ne j)}$
can be solved independently,
where the subscript $j$ denotes the number of grid points
in the $\varphi$-direction (see figure \ref{fig1}).
On the plane of $\varphi=\varphi_{(j)}$,
the initial position of $i$-th streamline
is set to be $(r,z)=(r_{\infty (i)},z_{\rm ini}$),
where $r_{\infty (i)}=(i-0.5)\Delta r$
(the subscript $i$ is the number of grid points in the $r$-direction).
The cross section at the initial point
is $\Delta S_{\infty (i)}=r_{\infty (i)} \Delta r \Delta \varphi$.

The optical depth of the streamlines
on the $\varphi=\varphi_{(j)}$ plane
is evaluated as follows.
The mass conservation law along the $i$-th streamline
is descried as
\begin{eqnarray}
  \rho_{(i)} v_{(i)} \Delta S_{(i)}
  = \rho_\infty v_\infty \Delta S_{\infty (i)},
  \label{masscons}
\end{eqnarray}
where the cross section perpendicular to the velocity vector of the streamline
is presented by $\Delta S_{(i)}$ (vertical cross section).
This is calculated as follows.
Assuming that $\Delta S_{r}$ is the cross-sectional area of the flow sliced in the $r$-direction (along the red line), the area projected onto the $z$-plane is $\Delta S_r \sin\theta_{(i)}=r_{(i)} l_{(i)}\Delta\varphi$. 
Here we have $\theta_{(i)}=\sin^{-1}(r_{(i)}/R_{(i)})$ where $\boldsymbol{R}_{(i)}$
is the position vector of the fluid element on the $i$-th streamline (see figure \ref{fig1}).
On the other hand, the relationship between $\Delta S_r$ and $\Delta S_{(i)}$ is $\Delta S_{(i)}=\Delta S_r\sin(\pi-\phi_{(i)})$.
Here, $\phi_{(i)}$ is defined as 
$\phi_{(i)}=\cos^{-1}|(\boldsymbol{v_{(i)}}/v_{(i)})
\cdot (\boldsymbol{R}_{(i)}/R_{(i)})|$.
From the above, $\Delta S_{(i)}$ is given by
\begin{eqnarray}
  \Delta S_{(i)}
  =\frac{r_{(i)} \ell_{(i)} \Delta\varphi}{\sin \theta_{(i)}}
  \sin \phi_{(i)}.
  \label{Sk}
\end{eqnarray}
Also, $\ell_{(i)}/\sin\theta_{(i)}$ means
the path length as photon (traveling in the $R$-direction)
passes through the streamline.
Note that 
the cross section of the streamline projected onto the $z$-plane
becomes a fan shape with the opening angle of $\Delta \varphi$,
since the coordinate value of $\varphi$
of the fluid element never changes.
Using equations (\ref{masscons}) and (\ref{Sk}),
the optical depth of the $i$-th streamline at the point of $\boldsymbol{R}_{(i)}$,
$\Delta \tau_{(i)}= \rho_{(i)} \kappa_{\rm dg} \ell_{(i)}/\sin\theta_{(i)}$
is rewritten as
\begin{eqnarray}
  \Delta\tau_{(i)}= \frac{\rho_\infty \kappa_{\rm dg} v_\infty r_{\infty (i)} \Delta r}
  {r_{(i)} v_{(i)} \sin\phi_{(i)}}.
  \label{dtau}
\end{eqnarray}

The optical depth, $\tau$, is basically estimated
by the sum of $\Delta \tau_{(i)}$ obtained above.
In concrete, we prepare $539$ straight lines radiating from the origin
on each planes where $\varphi$ is constant.
The polar angle of the straight lines is
$\theta_{(\mu)}= 0.005^\circ \mu$ for $\mu=1-240$
and
$\theta_{(\mu)}= 1.2^\circ + 0.6^\circ \mu$ for $\mu=241-539$
where $\theta$ is polar angle measured from the $z$-axis.
The subscript $\mu$ means the number of the straight lines.
Here, we define $\Delta\tau_{(\mu,i)}$
as the optical depth of the $i$-th streamline
at the intersection of the streamline
and the $\mu$-th straight line.
The optical depth of the $\mu$-th straight line
due to the 1st through $i$-th streamlines is evaluated as
\begin{eqnarray}
  \tau_{(\mu,i)}=\sum_{i'=1}^i \Delta\tau_{(\mu,i')}.
  \label{tau}
\end{eqnarray}

Applying equations (\ref{dtau}) and (\ref{tau}) to
the equation of motion (\ref{eq.motion2}),
we investigate in order from 1st streamline to $540$-th streamline
on the plane with $\varphi=\varphi_{(j)}$.
When we solve the $i$-th streamline, we employ $\Delta \tau_{(i)}$.
We also employ $\tau_{(\mu,i-1)}$ when the fluid element is
located between $m$-th and $(\mu+1)$-th straight lines.
Here, we set $\tau_{(\mu,i-1)}=0$ for $i=1$.
If the following conditions,
\begin{eqnarray}
  -v^z_{(i)} &<& \left( \frac{GM}{z_{(i)}} \right)^{1/2}
    \nonumber\\
  && \times
    \left\{ 1-\Gamma' e^{-\tau}\frac{1-e^{-\Delta\tau}}{\Delta\tau}
    \left[ \left(2\cos\psi -1\right) \delta+1 \right] \right\}^{1/2},
    \label{v_esc}
\end{eqnarray}
are satisfied when the streamline reaches the $z$-axis,
the gas in that streamline is regarded to accrete towards the central BH,
and $\rho_\infty v_\infty \Delta S_{\infty (i)}$ is added to
the mass accretion rate.
This criterion given by equation (\ref{v_esc}) is almost the same as that for \cite{hoyle1939}, but the right hand side is the escape velocity taking into account the radiation force. Since the $r$-component of the velocity becomes almost zero due to the collision when the gas arrives at the $z$-axis, the gas is considered to accrete when $-v^z$ is smaller than the escape velocity.
By repeating the above procedure
for 40 planes with different $\varphi$ ($j=1$-$40$),
the steady structure of the flows can be obtained.
The mass accretion rate, $\dot{M}$, is also revealed.

Finally we mention the treatment of streamline merging
since the streamlines sometimes intersect in our model.
When the $i$-th streamline intersects the $(i-1)$-th streamline,
we consider that the two streamlines merge at the intersection point.
That is, after the point, two streamlines becomes one streamline
and the fluid elements of $i$-th and $(i-1)$-th streamlines
move together.
After the intersection point,
the $(i-1)$-th streamline is cancelled
and we overwrite $\Delta\tau_{\mu,(i-1)}$ to zero.
The cross section at the starting point
of the single, merging stream line is set to be
$\Delta S_{\infty (i-1)} + \Delta S_{\infty (i)}$.
We consider that the mass is carried at the rate
with $\rho_\infty v_\infty (\Delta S_{\infty (i-1)} + \Delta S_{\infty (i)})$
along the merging stream line.
Based on the momentum conservation law,
the velocity immediately after the merging is given by
\begin{eqnarray}
  \boldsymbol{v}_{\rm merg}
  = \frac{\Delta S_{\infty (i-1)} \boldsymbol{v}_{(i-1)}
  +\Delta S_{\infty (i)} \boldsymbol{v}_{(i)}}
  {\Delta S_{\infty (i-1)} + \Delta S_{\infty (i)}}.
\end{eqnarray}
If the merged streamline intersects with $(i-2)$-th streamline,
two streamlines merge to form one streamline.
The same procedure as mentioned above is used
to modify the velocity and mass transport rate.
Such a process is repeated every time
intersection occurs thereafter.

\section{Result}
\label{Result}
\subsection{Structure of flow}
\label{Structure of flow}
\begin{figure}[t]
  \begin{center}
  \includegraphics[width=88mm]{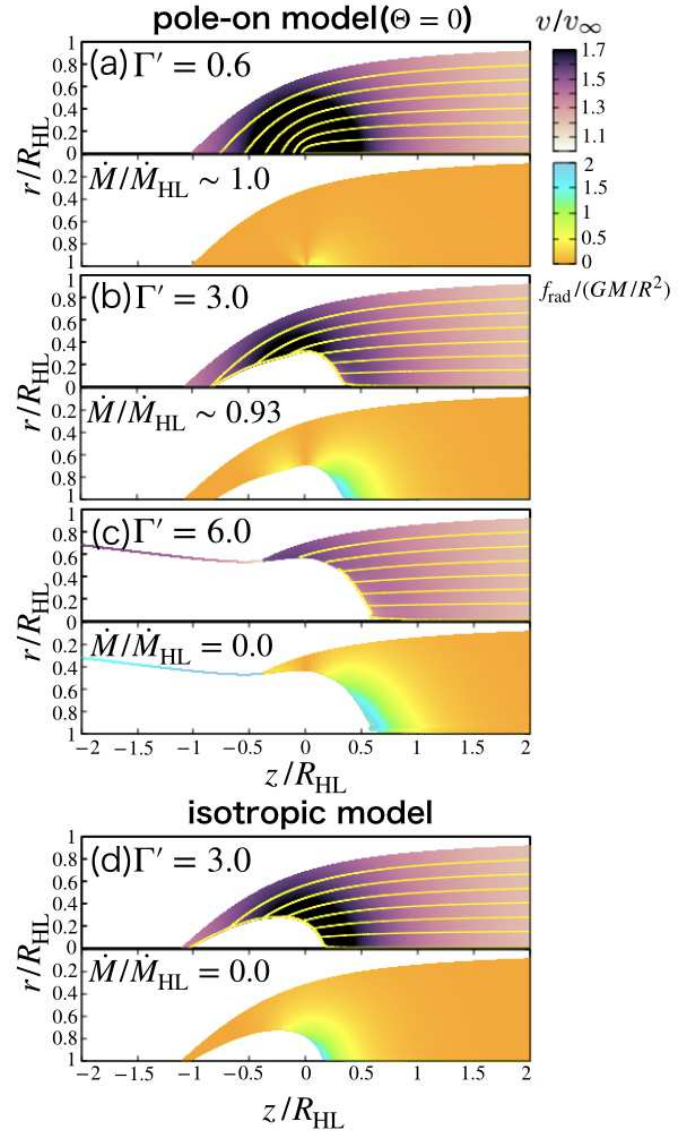}
  \end{center}
  \caption{
  The velocity field (upper color contour) 
  and the ratio of radiation force to gravity (lower color contour) 
  of the pole-on model, (panels a-c) 
  and isotropic model (panel d) 
  for the case of $\tau_{\rm HL}=3.3$.
  The yellow lines which overlaid in velocity distribution show the streamlines with 
  $i=1$,$70$,$150$,$230$,$310$,$390$,and $470$.
  We employ $\Gamma'=0.6$ (panel a), 3.0(panels b and d), and 6.0(panel c).
 }\label{fig2}
 \end{figure}

Figure \ref{fig2} shows
the velocity distribution 
and the ratio of radiation force to gravity
of the pole-on model, $\Theta=0$ (panels 2a-2c)
and the isotropic model (panel 2d)
for the case of $\tau_{\rm HL}=3.3$,
where $\tau_{\rm HL}$ is defined as 
$\tau_{\rm HL}= \rho_\infty \kappa_{\rm dg} R_{\rm HL}$.
Here the vertical and horizontal axes indicate 
$r/R_{\rm{HL}}$ and $z/R_{\rm{HL}}$, respectively.
On the top half of the panel,
the color contour indicates the velocity distribution 
and the yellow lines show the typical streamlines 
with $i=1$, $70$, $150$, $230$, $310$, $390$, and $470$.
The ratio of radiation force to gravity
is plotted on the bottom half of the panel.
We employ $\Gamma'=0.6$ (panel 2a),
$3.0$ (panels 2b and 2d), and $6.0$ (panel 2c).
In both models, the flow is axisymmetric
with respect to the $z$-axis.

In the sub-Eddington regime ($\Gamma'<1.0$),
the gravity dominates over radiation force
in the whole region.
This can be understood 
in the bottom half of the panel 2a.
It is found that 
the most of the area is orange 
and the yellow region appears near the central region.
Thus, the radiation force has 
little effect on the structure of the flow.
In the top half of the panel 2a,
we see that the streamline, which is nearly parallel to the 
$z$-axis at $z \gg R_{\rm HL}$,
is bent by the gravity of the BH
and reaches the $z$-axis behind the BH (in the region of $z<0$).
The velocity increases near the central region 
due to the gravity of the BH.
Most of the streamlines satisfy the accretion condition
(see equation (\ref{v_esc})),
so that the accretion rate 
becomes $\dot{M} \sim \dot{M}_{\rm HL} 
(\equiv \pi \rho_\infty v_\infty R_{\rm HL}^2 )$.

Even in the case of super-Eddington ($\Gamma'>1$),
the gas can accrete if the radiation force is weakened by attenuation.
Panel 2b shows that
the structure of the flow is not so different 
from that of the panel 2a,
except for the white region around the center.
The 470-th streamline (uppermost yellow line) 
is almost the same in panel 2a and panel 2b.
This means that the radiation force is too weak
to change the motion of the gas passing through 
the region far from the BH.
The reason for this is that
the radiation force is sufficiently weaker than gravity
due to the absorption by the dust.
Indeed, 
we can see in the bottom half of panel 2b
that the gravity is much larger than the radiation force
in the region where the 470-th flow passes through.

The trajectory of the gas approaching the center is bent 
and moves away from the $z$-axis.
For instance, 
the streamline with $i=1$ leaves the $z$-axis at $z/R_{\rm HL}\sim 0.3$.
This is because the strong radiation force pushes the gas outward.
In the bottom half of panel 2b,
the region that the radiation force exceeds the gravity
appears at $z\sim 0.3$-$0.5R_{\rm HL}$ and $r \lesssim 0.5R_{\rm HL}$ (light blue).
The 1st streamline passes around $r/R_{\rm HL}=0.3$ at $z=0$
while merging with other streamlines
and finally reaches the $z$-axis at $z/R_{\rm HL}\sim -0.75$.
Note that there is no streamlines in the white region around the center.
Although the structure of the flow is drastically affected 
by the radiation force around the central region,
most of the gas can accrete, 
and $\dot{M}/\dot{M}_{\rm HL}\sim 0.93$ 
even though $\Gamma'=3.0$. 
The streamlines that do not satisfy the accretion condition
are only a few streamlines that reach around $z\sim -1$.

The accretion is prevented if the radiation is too strong.
Panel 2c shows the results for the case of $\Gamma'=6.0$.
There are no streamlines which reach the $z$-axis.
In this case, the radiation force exceeds the gravity 
although the dust absorption reduces the radiation,
so that all the gas is blown away.
As shown in the panel,
it is found that, at the region of $z \lesssim -0.5$,
all stream lines merge and 
$f_{\rm rad}$ is larger than $GM/R^2$ on the streamline.

Panel 2d is the same as panel 2b but for isotropic model.
The overall structure is very similar to that 
of the pole-on model (panel 2b).
The many streamlines merge and 
reach $z$-axis at $z/R_{\rm HL}\lesssim -1.05$.
However, all of them do not satisfy the accretion condition ($\dot{M}=0$)
unlike in the case of the polo-on model,
in which most of the gas accretes ($\dot{M}/\dot{M}_{\rm HL} \sim 0.93$).
Such a significant difference in accretion rate 
is mainly caused by the position that 
the streamlines reach $z$-axis (reach point).
The streamlines in panel 2d reach the $z$-axis 
at a point slightly further from the center
than the streamlines in panel 2b.
Thus, the accretion condition is not satisfied unlike the case of the pole-on model,
and the accretion rate becomes zero.

Here we note that
the radiation flux (force) around the $z$-axis
tends to be stronger for the pole-on model
than for the isotropic model,
since the rotation axis of the disk coincides with the $z$-axis.
Compared with panel 2b and panel 2d,
we can see that the light blue region
near the $z$-axis is wider 
for the pole-on model than for isotropic model.
Nevertheless,
the accretion is feasible for the pole-on model
although the radiation prevent the accretion in the case of the isotropic model
(see appendix 1 for detailed reason).

\subsection{Accretion cross-section}
\begin{figure}[t]
  \begin{center}
   \includegraphics[width=80mm]{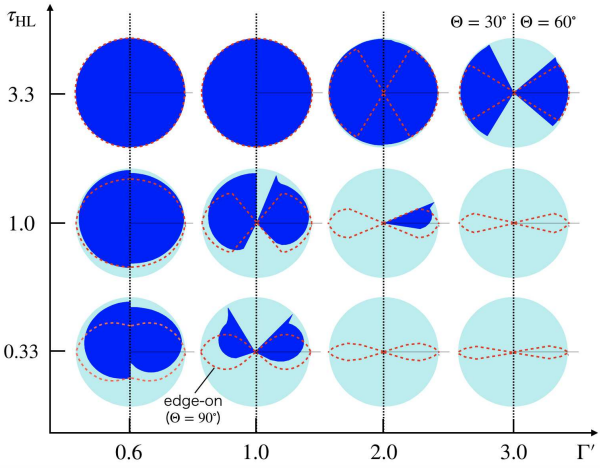}
  \end{center}
  \caption{
  Shapes of accretion cross section for various $\Gamma'$ and $\tau_{\rm HL}$.
  The red dotted line indicates the accretion radius for edge-on model ($\Theta=90^{\circ}$).
  The radius of light blue circle represents the  Hoyle-Lyttleton radii.
  The blue area is an accretion cross section 
  with $\Theta=30^{\circ}$ (left half) 
  and $\Theta=60^{\circ}$ (right half).
  }\label{fig3}
\end{figure}

The accretion radius is independent of $\varphi$
for the pole-on model and the isotropic model,
since the flow is axisymmetric with respect to the $z$-axis.
However, the accretion radius depends on $\varphi$ when $\Theta \ne 0$.
The red dotted line in figure \ref{fig3}
indicates the accretion radius for $\Theta=90^\circ$ (edge-on model).
The region enclosed by this line is called the accretion cross section, 
and streamlines that passes through the cross section at the plane of $z=z_{\rm ini}$ 
satisfies the accretion condition. 
Since the radius of the light blue circle in the figure \ref{fig3}
is $R_{\rm HL}$,
the normalized accretion rate, $\dot{M}/\dot{M}_{\rm HL}$,
is obtained from the ratio of area of the accretion cross section
to the area of the light blue circle.
The blue area is an accretion cross section with $\Theta=30^\circ$ 
(left half) and $60^\circ$ (right half).
Although we represent only half of the accretion cross sections
for $\Theta=30^\circ$ and $\Theta=60^\circ$, they are symmetric
with respect to the vertical lines.

\begin{figure*}[h]
  \begin{center}
   \includegraphics[width=150mm]{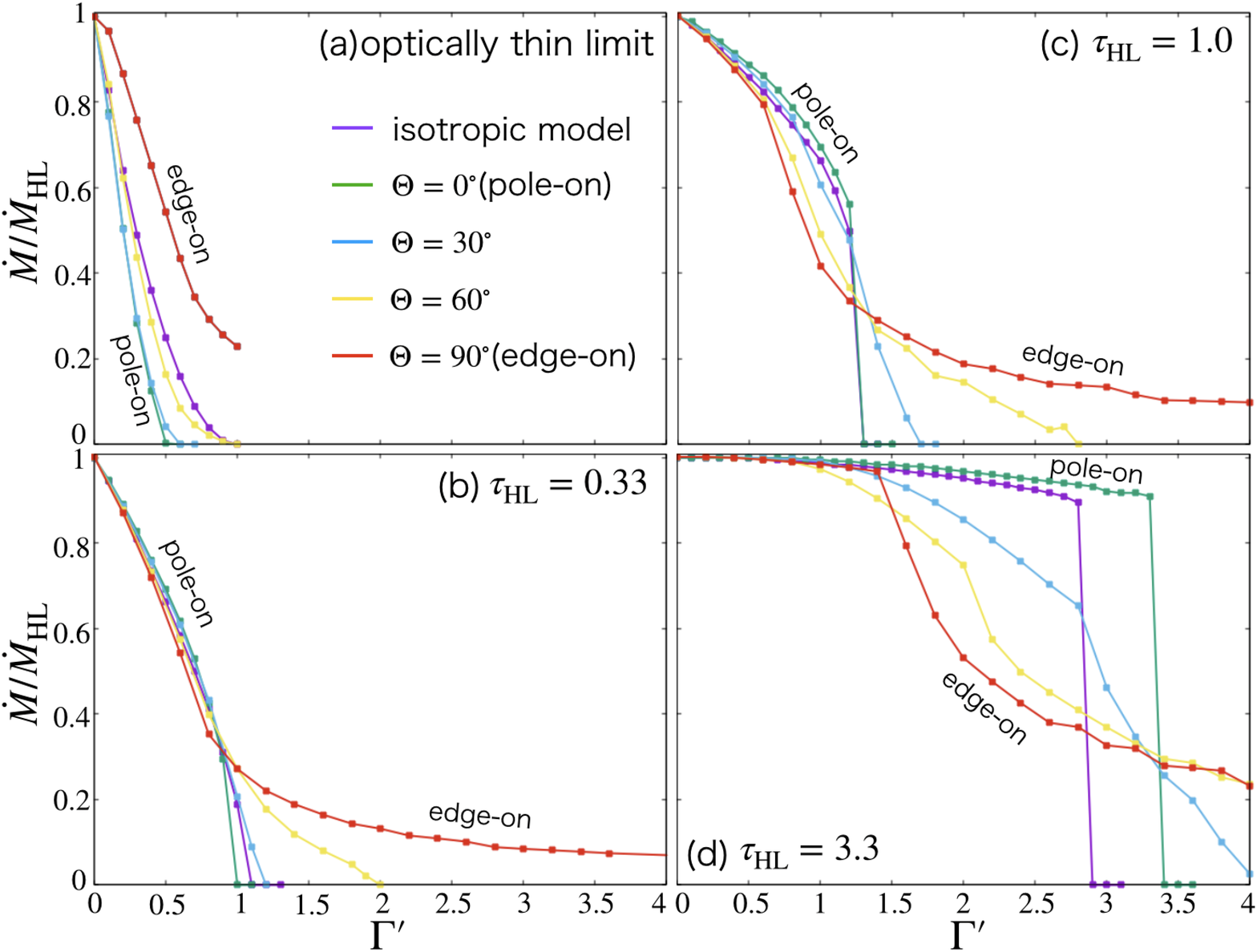}
  \end{center}
  \caption{
  Accretion rate normalized by  Hoyle-Lyttleton one as a function of $\Gamma'$ 
  for the case of 
  $\tau_{\rm{HL}}=0$ (optically thin limit, panel a), 
  $\tau_{\rm{HL}}=0.33$ (panel b), 
  $\tau_{\rm{HL}}=1.0$ (panel c),
  and $\tau_{\rm{HL}}=3.3$ (panel d). 
  Results for $\Theta=0^{\circ}$, $30$, $60$, and $90$ are shown by green, blue, yellow, and red lines, respectively. The purple line is for the isotropic model. 
  
  }\label{fig4}
 \end{figure*}

As shown in figure \ref{fig3},
the accretion cross section is larger in the upper left part of the figure for all cases.
This is because the smaller $\Gamma'$ is, 
and the larger $\tau_{\rm HL}$ is, 
the weaker the radiation force which works to prevent the gas accretion.
We find that, as it moves to the lower right side,
accretion becomes more difficult
via the strong radiation force.
For $(\Gamma', \tau_{\rm HL})=(2.0, 0.33), (3.0, 1.0)$, and $(3.0, 0.33)$,
the accretion are disappear for $\Theta=30^\circ$ and for $60^\circ$ 
via the strong radiation force.
We also find that 
the most difficult area to accrete 
is around $\varphi=0^\circ$ and $180^\circ$.
For instance, in the case of $\Theta=60^\circ$, 
the accretion cross section is broadened in the 
$\varphi=90^\circ$ direction, while 
the accretion radius is very small or zero
around $\varphi=0$ and $180^\circ$
for $(\Gamma', \tau_{\rm HL})=(1.0, 0.33), (1.0, 1.0), (2.0, 1.0), (3.0, 3.3)$.
This is because the radiation flux is large 
in the direction of the rotation axis of the disk.
On the contrary, 
around $\varphi=90^\circ$, 
which is closer to the disk plane, 
the radiation force is weaker, so accretion is easier.
Also, we find that the accretion cross section for $\Theta=30^\circ$ and $\Theta=60^\circ$ is not symmetrical to the plane of $\varphi=90^\circ$ in figure \ref{fig3} (see appendix 3 for detailed reason).
In particular,
for the edge-on model ($\Theta=90^\circ$),
the plane with $\varphi=90^\circ$ aligns with the disk plane, 
and thus the radiation flux is zero. 
Therefore, it is possible to accrete regardless of how large $\Gamma'$ becomes,
and the horizontally extended accretion cross section 
does not disappear in the lower right region of the figure.

\subsection{Accretion rate}
\label{accretion_rate}

Resulting mass accretion rate for $\tau_{\rm HL}=0$ (optically thin limit, panel \ref{fig4}a),
$0.33$ (panel \ref{fig4}b), $1.0$ (panel \ref{fig4}c), and $3.3$ (panel \ref{fig4}d)
is presented in figure \ref{fig4}
as a function of $\Gamma'$.
A purple solid line is for the isotropic model,
a green one for $\Theta=0^{\circ}$ (pole-on model),
a blue one for $\Theta=30^{\circ}$, an yellow one for $\Theta=60^{\circ}$,
and a red one for $\Theta=90^{\circ}$ (edge-on model).

In the optically thin limit,
the mass accretion rate decreases as an increase of $\Gamma'$ (see panel \ref{fig4}a).
The accretion rate in the pole-on model is the smallest of all models.
This is because the rotation axis of the disk in the pole-on model is aligned with the $z$-axis,
and the radiation force effectively acts on the gas flowing in from the $z$-direction.
For $\Gamma'>0.5$, the radiation force exceeds the gravity except in the region of $z\sim 0$
and the accretion rate becomes zero.
The radiation force in the region of $z\gg 0$ becomes weaker
as $\Theta$ approaches $90^\circ$,
hence the accretion rate in the edge-on model is larger than that in other cases.
It should be noted that the mass accretion rate in the edge-on model 
is not zero even if $\Gamma'$ is unity,
since the accretion through the vicinity of the disk plane
($\varphi=90^\circ$) is not prohibited (see appendix 1).
Our results for the pole-on and edge-on models 
are consistent with \citet{fukue1999}.

We can see in the panels \ref{fig4}b-\ref{fig4}d,
that the accretion rate increases as $\tau_{\rm HL}$ increases.
This is because the attenuation via the dust absorption weakens the radiation force.
This figure reveals that 
the accretion is possible in super-Eddington regime ($\Gamma'>1$) 
not only for edge-on model but also for all models
for $\tau_{\rm HL}\gtrsim 1$. 
Moreover, the dependence of the accretion rate 
on $\Theta$ also changes at the regime of $\tau_{\rm HL}\gtrsim 1$.
Although 
the accretion rate increases with increasing $\Theta$
for an optically thin limit as shown above,
such a trend is reversed as $\tau_{\rm HL}$ increases.
Indeed, at the regime of $\Gamma' \lesssim 1.0$ (panel \ref{fig4}b),
the accretion rate is approximately equal for all models when $\tau_{\rm HL}=0.33$
and increases as an increase of $\Theta$ 
in the case of $\tau_{\rm HL}=1.0$ (panel \ref{fig4}c).
This trend becomes more evident for $\tau_{\rm HL}=3.3$.
The panel \ref{fig4}d shows that the accretion rate tends to be larger 
as $\Theta$ is smaller except in the range of $\Gamma' \gtrsim 3.0$.

In the case of large optical thickness ($\tau_{\rm HL}=1.0, 3.3$),
the accretion rate for the pole-on model exceeds
that for the isotropic model as we have mentioned \S \ref{Structure of flow}.
The reason why that the accretion rate suddenly decreases to zero
is due to the merging of many streamlines (see appendix 1).
In addition, even when $\Gamma'$ is extremely large, 
the accretion rate of the edge-on model does not become zero.
This is because
the radiation force never exceeds the gravity
at around the disk plane (see appendix 1).

\section{Discussion}
\label{Discussion}
\subsection{Canonical Eddington ratio}

The accretion rate in our model is 
given by a function of the inclination angle of the accretion disk, $\Theta$,
the Eddington ratio for the dusty gas, $\Gamma'$, 
and the typical optical depth of the flow, $\tau_{\rm HL}$
(only $\Gamma'$ and $\tau_{\rm HL}$ for the isotropic model).
As shown in \S \ref{accretion_rate}, 
the accretion rate normalized by Hoyle-Lyttleton one 
is approximately expressed as
\begin{eqnarray}
  \frac{\dot{M}}{\dot{M}_{\rm{HL}}}=\min[f(\Gamma',\Theta,\tau_{\rm{HL}}),g(\Gamma',\Theta,\tau_{\rm{HL}})],
  \label{eq.dotM_1}
\end{eqnarray}
where $f$ and $g$ is given by 
\begin{eqnarray}
  f=
  \left\{
  \begin{array}{ll}
    \left(1+\frac{1}{4\tau_{\rm{HL}}}\Gamma'\right) \left(1- \frac{1}{2\tau_{\rm{HL}}^{1.5}}\Gamma'\right)
    & {\rm for\,\,\, isotropic} \\
    \left(1+\frac{1}{3\tau_{\rm{HL}}}\Gamma'\right) \left(1- \frac{1}{2\tau_{\rm{HL}}^{1.4}}\Gamma'\right)
    & {\rm for\,\,\, pole-on} \\
    \left(1+\frac{1}{8\tau_{\rm{HL}}}\Gamma'\right) \left(1- \frac{1}{2\tau_{\rm{HL}}^{1.2}}\Gamma'\right)
    & {\rm for\,\,\, edge-on} 
  \end{array}
  \right.,
\end{eqnarray}  
and
\begin{eqnarray}  
  g=
  \left\{
  \begin{array}{ll}
    q+1.8\tau_{\rm{HL}}^{1.6}(\tau_{\rm{HL}}-0.33) & {\rm for\,\,\, isotropic} \\
    q+1.8\tau_{\rm{HL}}^{1.95}(\tau_{\rm{HL}}-0.33) & {\rm for\,\,\, pole-on} \\
    0.5 \tau_{\rm{HL}}^{0.7}/\Gamma' & {\rm for\,\,\, edge-on} 
  \end{array}
  \right.
\end{eqnarray}
with $q=\left(1+2\tau_{\rm{HL}}\Gamma'\right) \left(1-\Gamma'\right)$.

On the other hand,
if the luminosity of the disk is given 
by $L=\eta \dot{M}_{\rm BH} c^2$
with $\eta$ being the energy conversion efficiency
and $\dot{M}_{\rm BH}$ being the accretion rate onto the BH,
then, $\dot{M}_{\rm BH}$ is related to $\Gamma'$ as
\begin{eqnarray}
  \frac{\dot{M}_{\rm BH}} {\dot{M}_{\rm{HL}}}
  =\frac{\Gamma'}{\dot{m}_{\rm dg}},
  \label{eq.dotM_2}
\end{eqnarray}
where 
\begin{eqnarray}
  \dot{m}_{\rm dg}
  =\frac{\kappa_{\rm dg}}{\kappa_{\rm es}}\frac{\eta c^2 \dot{M}_{\rm HL}}{L_{\rm Edd}}
  =\frac{\eta c\tau_{\rm{HL}}}{2v_{\infty}}.
\end{eqnarray}

Solving for $\Gamma'$, assuming $\dot{M}$ and $\dot{M}_{\rm BH}$ are equal, 
we get a canonical Eddington ratio, $\Gamma_{\rm{can}}'$.
This is the ratio that is realized 
when the accretion rate at the Hoyle-Lyttleton mechanism which takes into account the effect of radiation, 
and the accretion rate from the disk to the BH, 
are consistent.
The canonical Eddington ratio, $\Gamma'_{\rm can}$, is given by
\begin{eqnarray}
  \Gamma_{\rm{can}}'=\min [h(\dot{m}_{\rm{dg}},\tau_{\rm{HL}}),k(\dot{m}_{\rm{dg}},\tau_{\rm{HL}})].
\end{eqnarray}
Here, $h$ is 
\begin{eqnarray}  
  h= \left[ \sqrt{ \xi^2 +\frac{\dot{m}_{\rm{dg}}^2} {\gamma \tau_{\rm{HL}}^{\zeta}} } -\xi \right]
      \left( \frac{\dot{m}_{\rm{dg}}} {2\gamma \tau_{\rm{HL}}^{\zeta}} \right)^{-1},
\end{eqnarray}
where
\begin{eqnarray}  
  \xi= 
  \left\{
    \begin{array}{ll}
      \frac{\dot{m}_{\rm{dg}}} {2\tau_{\rm{HL}}^{1.5}}
      -\frac{\dot{m}_{\rm{dg}}} {4\tau_{\rm{HL}}}+1
      & {\rm for\,\,\, isotropic} \\
      \frac{\dot{m}_{\rm{dg}}} {2\tau_{\rm{HL}}^{1.4}}
      -\frac{\dot{m}_{\rm{dg}}} {3\tau_{\rm{HL}}} +1
      & {\rm for\,\,\, pole-on} \\
      \frac{\dot{m}_{\rm{dg}}} {3\tau_{\rm{HL}}^{1.2}}
      -\frac{\dot{m}_{\rm{dg}}} {8\tau_{\rm{HL}}} +1
      & {\rm for\,\,\, edge-on} 
    \end{array}
  \right.,
\end{eqnarray}
with $(\gamma, \zeta)$ being $(2, 2.5)$ for the isotropic model,
$(1.5, 2.4)$ for the pole-on model,
and $(6, 2.2)$ for the edge-on model. 
Also, $k$ is given by
\begin{eqnarray}  
  k=
  \left\{
  \begin{array}{ll}
    \left(4\tau_{\rm{HL}} \dot{m}_{\rm{dg}} \right)^{-1}
    \left[ \sqrt{p^2+s_1}-p \right]
    & {\rm for\,\,\, isotropic} \\
    \\
    \left(4\tau_{\rm{HL}} \dot{m}_{\rm{dg}} \right)^{-1}
    \left[ \sqrt{p^2+s_2}-p \right]
    & {\rm for\,\,\, pole-on} \\
    \\
    \sqrt{
    0.5 \dot{m}_{\rm{dg}} \tau_{\rm{HL}}^{0.7}
    }
    & {\rm for\,\,\, edge-on} 
  \end{array}
  \right.,
\end{eqnarray}
where $p=\dot{m}_{\rm{dg}}-2\dot{m}_{\rm{dg}} \tau_{\rm{HL}}+1$,
and 
\begin{eqnarray}
  s_1&=&8\tau_{\rm{HL}} \dot{m}_{\rm{dg}}^2 \left[1+1.8\tau_{\rm{HL}}^{1.6}
  \left(\tau_{\rm{HL}}-0.33\right) \right] \\
  s_2&=&8\tau_{\rm{HL}} \dot{m}_{\rm{dg}}^2 \left[1+1.8\tau_{\rm{HL}}^{1.95}
  \left(\tau_{\rm{HL}}-0.33\right) \right] .
\end{eqnarray}

\begin{figure}[t]
 \begin{center}
  \includegraphics[width=85mm]{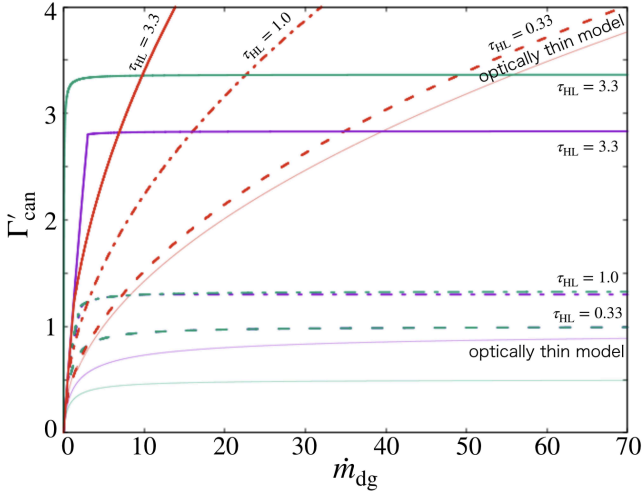}
 \end{center}
 \caption{
 Canonical Eddington ratio, $\Gamma'_{\rm can}$, as function of $\dot{m}_{\rm dg} (=\eta c \tau_{\rm HL}/(2v_{\infty}))$.
 The purple line is for the isotropic model,
 the green line is for the pole-on model $(\Theta=0^{\circ})$,
 and the red line is for the edge-on model$(\Theta=90^{\circ})$.
 The thick solid, dot-dashed, and dashed lines indicate for $\tau_{\rm{HL}}=3.3$, $1.0$, and $0.33$. The result for the optically thin limit is shown by the thin solid line. 
 
 }\label{fig5}
\end{figure}

Figure \ref{fig5} shows $\Gamma'_{\rm{can}}$ as a function of $\dot{m}_{\rm{dg}}$.
We find that $\Gamma'_{\rm can}$ increases with an increase of $\dot{m}_{\rm dg}$
and is almost constant in the region of $\dot{m}_{\rm dg} \gtrsim 10$ 
for the isotropic and pole-on models.
The maximum values of $\Gamma'_{\rm can}$ in figure \ref{fig5}
corresponds to the upper limits of $\Gamma'$ shown in figure \ref{fig4}.
This is because there is no canonical Eddington ratio 
that exceeds the upper limit,
since the radiation force prevents the gas accretion.
For instance,
the maximum value of $\Gamma'_{\rm can}$ in the pole-on model is $3.4$, 
which is consistent with the upper limit of $\Gamma'$
in the panel d in figure \ref{fig4}.

On the other hand, 
in the edge-on model, there is no upper limit for $\Gamma'$
(see figure \ref{fig4}).
Hence, $\Gamma'_{\rm can}$ increases as $\dot{m}_{\rm dg}$ increases.
This is because, even though the disk luminosity is large, 
the gas can accrete through the region near the disk plane, 
$\varphi=90^\circ$ (see figure \ref{fig3}).

\subsection{Evolution of Luminosity}

In this section, 
we first discuss the evolution of a BH 
with  
an accretion disk 
as it moves through a high density region.
If the initial luminosity of the accretion disk is different from the canonical luminosity, $\kappa_{\rm dg}\Gamma'_{\rm can} L_{\rm Edd}/\kappa_{\rm es}$, the luminosity gets close to the canonical one and should become same value finally.
This final state corresponds to that the accretion rate of the disk onto a BH becomes consistent with the one of the Hoyle-Lyttleton mechanism which takes into account the effect of radiation (i.e., the accretion rate from a larger scale to the disk).

The timescale 
of which the initial $\Gamma'$ matches $\Gamma'_{\rm can}$ is
estimated by 
using the solution of the standard disk model 
(\cite{kato1998}) as follows,
\begin{eqnarray}
  t_{\rm{vis}}
  &\sim & 3\times10^2
  \left( \frac{\alpha_{\rm{vis}}}{0.1}\right)^{-\frac{4}{5}}
    \left(\frac{M}{100M_{\odot}}\right)^{-\frac{9}{10}}
  \left(\frac{v_{\infty}}{100\,\rm{km\;s^{-1}}}\right)^{-\frac{8}{5}}
\nonumber\\
  &\ &\times\left(\frac{n_{\infty}}{10^5\, \rm{cm^{-3}}}\right)^{-\frac{3}{10}}
  \left( \frac{\chi}{0.1}\right)^{\frac{5}{2}}
  \;\rm{yr},
\end{eqnarray}

where $\alpha_{\rm{vis}}$ is the viscosity parameter.
Here we assume that 
the typical angular momentum of the gas
accreting onto the center is $\chi v_\infty R_{\rm HL}$,
where $\chi$ is constant less than $1$.
The angular momentum of the accreting gas
would depend on the anisotropy of the accretion cross section 
(see figure \ref{fig3}) 
and/or the nonuniform distribution of the gas density.

On the other hand, 
the timescale of the dynamical friction is estimated as
(\cite{chandrasekhar1943}; \cite{binney1987})
\begin{eqnarray}
  t_{\rm{fric}}
  &\sim& 8 \times10^8
  \left(\frac{v_\infty}{100\,\rm{km\;s^{-1}}}\right)^3
  \left(\frac{M}{100M_{\odot}}\right)^{-1}
  \nonumber\\
  &\ &\times \left(\frac{n_{\infty}}{10^5\,\rm{cm^{-3}}}\right)^{-1}
  \;\rm{yr},
\end{eqnarray}

where the Coulomb logarithm is supposed to be $10$.
Hence, $t_{\rm vis}$ is smaller than $t_{\rm fric}$
if 
\begin{eqnarray}
  \left(\frac{\alpha_{\rm{vis}}}{0.1}\right)^{4/5}&&
  \left(\frac{n_{\infty}}{10^5\,\rm{cm^{-3}}}\right)^{-7/10}
  \left(\frac{v_{\infty}}{100\,\rm{km\;s^{-1}}}\right)^{23/5}\\ \nonumber
  &\times&
  \left(\frac{M}{100M_{\odot}}\right)^{-1/10}
  \left( \frac{\chi}{0.1}\right)^{-5/2}
  >
  3 \times 10^{-7}.
\end{eqnarray}

This is the case,
the disk luminosity varies to the canonical luminosity 
($\Gamma'$ becomes $\Gamma'_{\rm can}$), 
and then the velocity of BH decreases gradually
via the dynamical friction.
If $t_{\rm vis}>t_{\rm fric}$, 
the initial $\Gamma'$ is kept almost constant while the velocity of a BH changes significantly. 
Detailed study of the dynamical friction of the moving BH
by the hydrodynamics simulations was recently attempted 
by \citet{li2019}.
They show that the compact object decelerates due to dynamical friction when the gravity of all the gas in a large simulation region is taken into account, but acceleration would occur when the gravity of the gas of a few Bondi radius is taken into account.
In addition, \citet{toyouchi2020} showed 
by the Radiation hydrodynamics simulations 
that the moving BH could be accelerated.

Here we note that 
$\Theta$ also changes on the timescale of $t_{\rm vis}$.
The direction of the angular momentum of the gas 
passing through the accretion cross section, $v_\infty r_\infty$,
is perpendicular to the $z$-axis.
Therefore, the rotation axis of the accretion disk 
formed by the accretion gas via the Hoyle-Lyttleton mechanism which takes into account the effect of radiation
is also perpendicular to the $z$-axis.
Edge-on accretion ($\Theta=90^\circ$) would be realized after $t_{\rm vis}$,
and the disk luminosity is thought to be the canonical luminosity 
for $\Theta=90^\circ$.
For example,
in the case of $M=500~ M_\odot$, $v_{\infty}=100~ \rm{km\;s^{-1}}$, 
and $n_\infty=10^5\,{\rm cm^{-3}}$,
we have $\tau_{\rm{HL}} \sim 0.33$ and $\dot{m}_{\rm{dg}}\sim 50$.
The canonical Eddington ratio is $\Gamma'_{\rm{can}}\sim 3.4$.
Since the disk luminosity is estimated to be 
$L \sim 10^{38} (M/500~M_\odot)[(\kappa_{\rm dg}/\kappa_{\rm es})/1900]~
\rm{erg\;s^{-1}}$,
and since the ultraviolet/optical photons emitted from the disk 
is effectively absorbed and infrared photons are reemitted,
the intermediate mass BHs moving in high density gas  
can be identified as bright sources in the infrared band.
Also, the X-ray and radio emission of the disk,
which transmit through the dusty-gas, would be observed.
The detailed spectra should be obtained 
by the multi-wavelength radiation transfer calculations.

Throughout the present study, we assumed that the dust exists everywhere in the gas flow.
If we roughly evaluate the temperature of the dust,
$T_{\rm dust}$, from the relation of 
$\sigma T_{\rm dust}^4 \sim  \Gamma'  
\left( \kappa_{\rm es}/\kappa_{\rm dg} \right) 
 \left( L_{\rm E}  /4 \pi R^2 \right)$
with $\sigma$ being the Stefan-Boltzmann constant,
the ratio of the sublimation radius of the dust,
$R_{\rm subl}$, to the Hoyle-Lyttleton radius is given by 
\begin{eqnarray}\label{subl_radi}
\frac{R_{\rm subl}}{R_{\rm HL}}
\sim
6 &\times& 10^{-3}~ \Gamma'^{1/2}
\left( \frac {M_{\rm BH}}{100M_{\odot}} \right)^{-1/2}
\left( \frac {v_{\infty}}{100\;\rm km\;s^{-1}} \right)^{2}
\\ \nonumber
&\times&
\left( \frac {T_{\rm dust}}{1500\;\rm K} \right)^{-2}
\left( \frac {\kappa_{\rm dg}/\kappa_{\rm es}}{1900} \right)^{-1/2}.
\end{eqnarray}

This relation implies that 
the region where the dust is sublimated 
is negligibly small
and our treatment in this paper, 
which does not consider the effect of sublimation of the dust, is reasonable.

\subsection{Future work}
\label{Future work}
By taking account of the dilution of the radiation
via the dust absorption,
we study the steady structure around the moving BHs
surrounded by the luminous accretion disks.
Although the hydrodynamic effects are neglected in the present work,
they may play an important role.
One of them is the shock wave.
Bow shock is thought to be formed 
if the speed of the moving BH,
which corresponds to the velocity of the gas flow ($v_\infty$)
in the BH rest frame, 
is greater than the sound speed.
Indeed, the occurrence of the bow shock was reported by 
the hydrodynamics simulations (\cite{hunt1971};  \cite{hunt1979}; \cite{shima1985}; \cite{ruffert1994}; \cite{ruffert1994}).
Our results also show that the streamlines are bending and merging. 
Thus shock waves can be generated by the collision of the gas flows. 
The velocity as well as the density of the flow changes at the shock surface, 
causing the pressure gradient force might become negligible. 
For instance, the motion of the gas in the $\varphi$-direction might be induced, 
although the gas moves in a $\varphi$-constant plane in the present model. 
The change in the density distribution causes the change in the radiation field. 
Then, the gas flow should also change. 
In order to reveal the structure of the flow by taking account of the shock, we need to perform multidimensional radiation hydrodynamics simulations. 
On the other hand, the accretion rate may increase with the occurrence of shock waves. 
This is because the release of thermal energy via the emitting photons reduces the total energy of the gas, 
making it more likely to accrete than in the absence of the shock. 
Also, if the dust is destructed, 
the radiative force will be ineffective and the gas accretes more easily. 
In order to understand the dust destruction process, 
it is necessary to study the structure of the shock wave in detail. 
This requires radiation hydrodynamics simulations, which is an important future work.

In the present paper, 
although we consider the radiation from the accretion disk around the BH,
the jets and disk winds might be launched from the disk (\cite{li2019}).
Such outflows also change the structure of the flow around the moving BHs
through the collision between the out-flowing matter and the interstellar gas.
Research for the effects of the outflows 
are left as an important future work.
In addition, 
although we focus on the steady structures in this work,
the time variations might occur.
As the disk luminosity (accretion rate) increases, 
the enhanced radiation force works to reduce the accretion rate.
Then the disk luminosity decreases via the reduction of the accretion rate,
leading to weakening of radiative force
and increasing the accretion rate.
Thus, the increase and decrease in the disk luminosity (accretion rate)
are repeated.
The time variation of the luminosity 
might cause the time-dependent outflows.
In order to investigate Bondi-Hoyle-Lyttleton accretion 
including the effects mentioned above,
we plan to perform multi-dimensional radiation hydrodynamics simulations.

In the present work, 
we consider a situation that 
the absorption coefficient of the dust 
is larger than the electron scattering opacity.
The absorption coefficient of dusty-gas 
depends on the dust-to-gas mass ratio,
which is thought to relate to the star formation history.
The dust-to-gas mass ratio in our galaxy is estimated to be about 0.01, 
but it would be much larger in quasars or starburst galaxies (\cite{solomon1997}; \cite{watson2015}). 
In contrast, it is predicted to be extremely small 
in the very early universe (\cite{inoue2016}).
In addition, the absorption coefficient of dusty-gas 
also depends on $\alpha$, 
which is determined by the size distribution 
and chemical composition of the dust grain (\cite{draine1984}).
It also depends on the spectrum of the disk emission.
The value of $\alpha$ tends to be larger 
if the disk shines mainly in UV/optical band.
In contrast, $\alpha$ becomes small
if the radiation is dominant in the radio wave or x-ray band.
In the case of the standard disk around the stellar mass BHs/intermediate mass BHs, 
as the BH mass is smaller and the Eddington ratio is larger, 
the temperature of the disk is known to increase 
and the X-ray emission is enhanced.
Meanwhile, 
the radiatively inefficient accretion flows,
which appears in the regime where the accretion rate is very small ($\ll L_{\rm E}/c^2$),
is thought to mainly emit photons at the radio band.
The study taking account of spectra of the accretion disks
is also left as an important future work.
Moreover,
the detailed calculations of observable spectra, 
consisting of penetrating intrinsic radiation from the disk and the IR photons from the dust, 
would be important for comparison with observational data.
We will investigate the above issues in future work.

\section{Conclusions}
\label{Conclusion}

In this study, 
we have investigate the steady structure of Hoyle-Lyttleton accretion of dusty-gas, 
assuming that a BH with an accretion disk moves in dense gas.
For this purpose, we solve the equation of motion 
which incorporate the gravity of the central object and the radiation force due to dust absorption,
by coupling with the continuity equation and attenuation of the radiation.
Here, the scattering process of dust is assumed to be negligibly small.

We find that the gas accretion is possible 
even when the luminosity is larger than the Eddington luminosity for the dusty-gas 
if the flow is so optically thick that radiation is sufficiently weakened by the dust absorption.
For example, in the case of $\tau_{\rm HL}=3.3$ and $\Theta=0$ (pole-on model), 
the accretion rate is 92\% of Hoyle-Lyttleton one even at $\Gamma'=3.0$.
Also, in the edge-on model ($\Theta=90^\circ$), 
the gas accretes at the rate of 32\% of Hoyle-Lyttleton rate.
Note that in the regime of $\Gamma'$, 
where the gas accretion is possible for all models, 
the accretion rate tends to be greater as $\Theta$ is smaller
for the case of $\tau_{\rm HL} \gtrsim 1.0$.
This is the opposite trend to the optically thin case.
As mention in Appendix 1, 
this is due to the effects of optical thickness 
and anisotropic radiation field of the disk.

If the optical depth is not enough 
or if the luminosity is extremely large,
the radiation force is not sufficiently weakened
so that the accretion rate is reduced or the gas accretion becomes impossible.
This is confirmed that 
both the accretion rate for $(\Gamma'$, $\tau_{\rm HL})=(3.0, 1.0)$
and for $(\Gamma'$, $\tau_{\rm HL})=(3.4, 3.3)$
is smaller than that for $(\Gamma'$, $\tau_{\rm HL})=(3.0, 3.3)$.
For $(\Gamma'$, $\tau_{\rm HL})=(3.0, 1.0)$,
the accretion rate is zero in the case of $\Theta=0^\circ$ 
and 13\% of the Hoyle-Lyttleton one in the case of $\Theta=90^\circ$.
For $(\Gamma'$, $\tau_{\rm HL})=(3.4, 3.3)$,
the accretion rate zero when $\Theta=0^\circ$ 
and 30\% of the Hoyle-Lyttleton one when $\Theta=90^\circ$.
These values are smaller than those for for $(\Gamma'$, $\tau_{\rm HL})=(3.0, 3.3)$.

Although the accretion rate rapidly goes to zero 
as $\Gamma'$ becomes large for the case of $\Theta=0^\circ$,
the decrease in the accretion rate due to the increase in $\Gamma'$ 
becomes moderate as $\Theta$ approaches $90^\circ$.
This is because even if the luminosity of the disk is very large, 
the radiative flux in the direction of the disk plane is small, 
so the radiation force is less effective in preventing accretion 
of the gas passing around the disk plane.
Hence, a sharp decrease does not appear for models with $\Theta \gtrsim 60^\circ$.
In particular, in the case of $\Theta=90^\circ$, 
the accretion rate does not decrease to zero even if $\Gamma'$ is extremely large.

Furthermore, we obtain the canonical Eddington ratio,
where the mass accretion rate and the disk luminosity become consistent 
(the accretion rate by Hoyle-Lyttleton mechanism which takes into account the effect of radiation and the disk luminosity satisfy the relation of $L=\eta \dot{M} c^2$), 
with using fitting functions for the numerical results.
The Eddington ratio of a BH accretion disk moving in dense dusty-gas 
is likely to evolve into a canonical Eddington ratio.
The rotation axis of the disk also change and would be $\Theta=90^\circ$ at that time.
For instance, 
if a BH with $500M_\odot$ moves in a dense dusty-gas with $10^5\,{\rm cm^{-3}}$ at the speed of $100\rm{km\;s^{-1}}$,
the canonical Eddington ratio is $\sim 3.4$.
Since the disk luminosity is $L \sim 10^{38} (M/500M_\odot)[(\kappa_{\rm dg}/\kappa_{\rm es})/1900]~\rm{erg\;s^{-1}}$, 
intermediate-mass BHs moving in dense dusty-gas 
would be observed as bright sources in the infrared band.
Although there is some uncertainty about the absorption coefficient of the dust,
which is thought to depend on the dust-to-gas mass ratio, 
the composition and the size distribution of the dust grain, 
and the spectrum of the accretion disk.

\section{Acknowledgement}
This research used computational resources in Center for Computational Sciences, University of Tsukuba.
This work was supported by JSPS KAKENHI Grant Numbers JP18K03710 (KO), 
17H04827, 20H04724 (HY), National Astronomical Observatory of Japan (NAOJ) 
ALMA Scietific Research Grant Number 2019-11A (HY)
and by MEXT as “Program for Promoting Researches on the Supercomputer Fugaku” 
(Toward a unified view of the universe: from large scale structures to planets, KO) 
and by Joint Institute for Computational Fundamental Science (JICFuS, KO).

\appendix
\section{Effects of disk emission}
\label{Effects of disk emission}

\begin{figure}[t]
  \begin{center}
   \includegraphics[width=80mm]{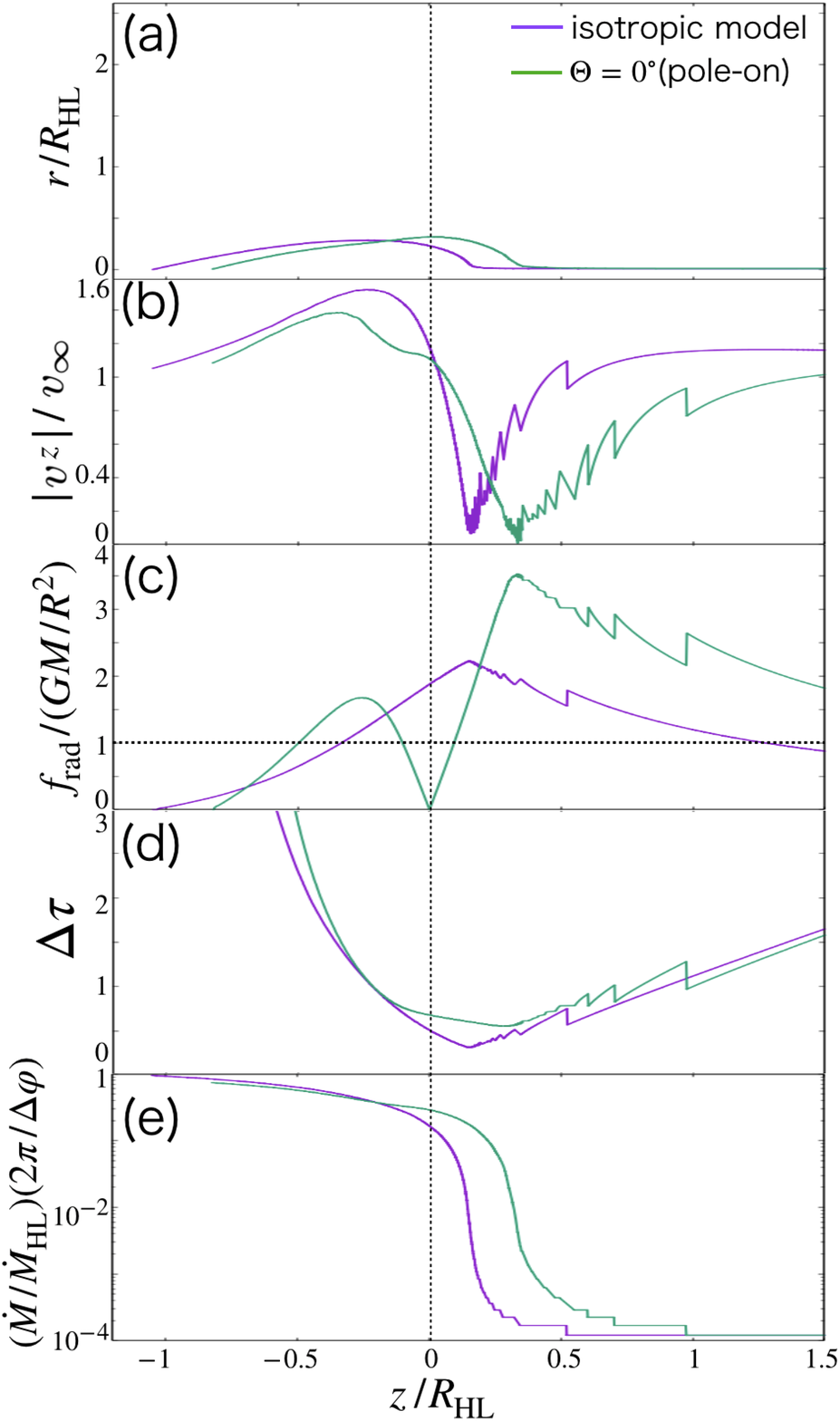}
  \end{center}
  \caption{
   The streamlines closest to the $z$-axis (a) for the case of $\tau_{\rm HL}=3.3$ and $\Gamma'=3.0$. The purple and green lines are for the isotropic model and for the pole-on model. We also plot the $z$-component of the velocity (b), the ratio of radiation force to gravity (c), the optical depth (d) on the streamlines. The mass transport rate is shown in panel e. 
  }\label{fig6}
 \end{figure}

Panel \ref{fig6}a in figure \ref{fig6} shows the streamlines
closest to the $z$-axis in the case of $\Gamma'=3.0$
for the pole-on model (purple) and 
for the isotropic model (green).
These lines are the same as those shown in panels 2b and 2d of figure \ref{fig2}.
We also plot the $z$-component of the velocity (panel \ref{fig6}b),
the radiation force (panel \ref{fig6}c), 
and the optical depth (panel \ref{fig6}d) on the streamlines. 
The panel 3e indicates the mass 
transported along the streamline per unit time (mass transport rate).
The mass transport rate increases by the merging of the streamlines.

As we have mentioned in \S \ref{Structure of flow},
the streamline is approximately parallel to the $z$-axis at $z/R_{\rm HL} \gg 1.0$ 
for both models (panel \ref{fig6}a).
The streamlines bend significantly in the region where $z/R_{\rm HL} \lesssim 0.3$.
The gas leaves the $z$-axis once, 
approaches it again, and finally reaches the $z$-axis.
In panel \ref{fig6}b,
we can see that $|-v^z|$ substantially decreases
in the region of $0.25\lesssim z \lesssim 1.0$ for both models.
In this region, the radiation force is stronger than the gravity (see panel \ref{fig6}c),
leading to a slow down.

Here we note that a jagged structure as shown in panels \ref{fig6}b-\ref{fig6}e
is caused by the merging of the streamlines.
In our method, when two streamlines merge into one, 
the mass transport rate and the velocity change
based on the mass conservation law and the momentum conservation law
(see \S \ref{Numerical Method}).
The mass transport rate becomes 
the sum of the mass transport rate of the two streamlines.
Thus, the velocity discontinuously changes,
producing the jagged structure in panel \ref{fig6}b.
By repeating the merger,
the mass transport rate 
increases from right to left in panel \ref{fig6}e.
Furthermore,
by the change of the velocity and the mass transport rate,
the jagged structure appears in panels \ref{fig6}c and \ref{fig6}d.

As shown in panel \ref{fig6}b, 
the velocity ($|-v^z|$) of the isotropic model
increases in the region of $0\leq z\leq 0.15$,
even though the radiation force is stronger than the gravity.
This is caused by the the merging of the streamlines.
The streamlines farther from the $z$-axis (streamlines with large $i$)
tends not be decelerated by the radiation force,
since the incident radiation is more attenuated by the dust absorption.
Such less decelerated streamlines merge into the streamline with $i=1$ 
so that the velocity increases.
The velocity, $-v^z/v_\infty$, which is $\sim 0.1$ at $z/R_{\rm HL} \sim 0.15$,
becomes $\sim 1.0$ at around $z/R_{\rm HL}=0$.
The increase in the velocity due to merging continues
until $z/R_{\rm HL}\sim -0.2$.

Similar to the isotropic model,
the pole-on model also shows an increase of the velocity due to merging.
The velocity, $|-v^z|/v_\infty$, which is $\sim 0$ at $z/R_{\rm HL} \sim 0.3$,
becomes $\sim 1.4$ at around $z/R_{\rm HL} = -0.3$.
However, the effect of the unisotropic radiation appears 
in the case of the pole-on model.
The radiation force is less than the gravity around $z=0$.
The reason is that the plane of $z=0$ coincides with 
the disk plane.
Thus, the streamline begins to approach the $z$-axis 
before that of the isotropic model does.
As a result,
The upper and lower positions of the streamlines are 
switched at around $z\sim -0.2$ (panel \ref{fig6}a).

It is found that 
the radiation force is less than gravity
in the region of $z\lesssim -0.5$ (pole-on model)
and $z\lesssim -0.3$ (isotopic model).
The reduction of the radiation force is 
induced by the attenuation of the radiation.
In the above regions,
the optical depth of the stream line 
is much larger than unity (panel \ref{fig6}d),
since many streamlines merge into the streamline,
and since the flow is concentrated in the $z$-axis
(the streamlines approach to the $z$-axis).
Hence, the radiation force cannot increase the velocity,
and the gas reaches the $z$-axis with a slight deceleration.
Finally, the gas accretes onto the center 
for the case of the pole-on model.
However, in contrast, the stream line 
for the isotropic model does not satisfy the accretion condition
since reach point is slightly farther from the center 
in comparison with the pole-on model.

To sum up,
(i)strong radiation force initially leads to the deceleration and merging of the streamlines
(ii)The streamlines approach the $z$-axis 
due to weak radiation force at around $z=0$ 
(the disk plane).
(iii)The radiation force cannot effectively accelerate the gas 
behind the BH.
Therefore, the accretion rate becomes large for the pole-on model.
For the case of the isotropic model, (ii) does not occur 
so that the accretion rate is drastically reduced.
In order to achieve a large accretion rate, 
both the unisotropic radiation fields 
and the attenuation of the radiation are needed.


\begin{figure}[t]
  \begin{center}
   \includegraphics[width=82mm]{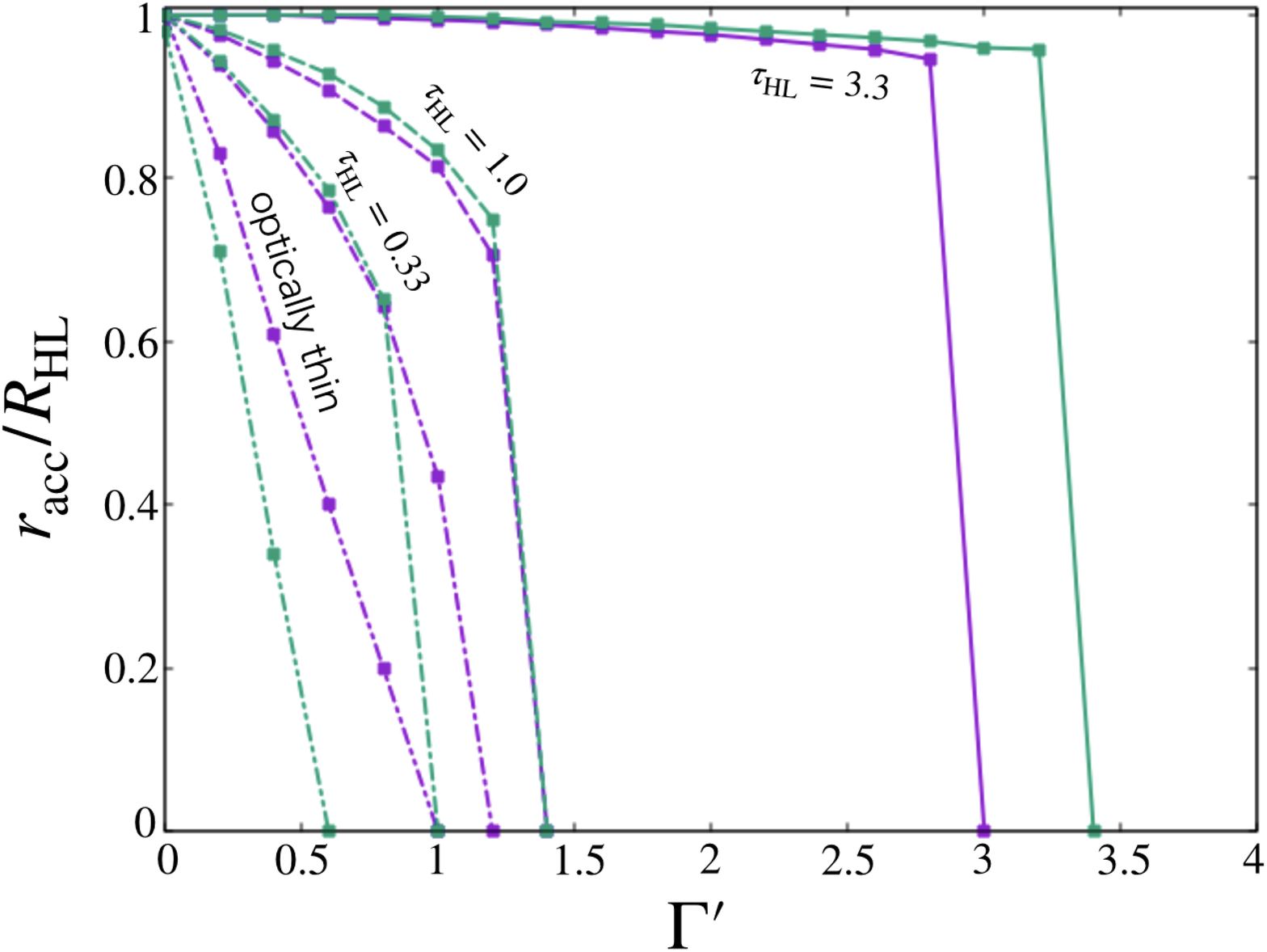}
  \end{center}
  \caption{
  Accretion radius normalized by the Hoyle-Lyttleton one,
  $r_{\rm{acc}}/R_{\rm{HL}}$,
  as a function of the effective Eddington ratio $\Gamma'$.
  The results for the isotropic and pole-on models are plotted by purple and green lines. 
  The solid, dashed, and dash-dotted lines are for $\tau_{\rm{HL}}=3.3$, $1.0$ and $0.33$, respectively. The dashed-and double-dotted line indicates for the case of optically thin limit. 
  
  }\label{fig7}
\end{figure}

Figure \ref{fig7} shows the accretion radius $r_{\rm acc}$
normalized by the Hoyle-Lyttleton accretion radius $R_{\rm acc}$
as a function of $\Gamma'$ for the isotropic model 
and the pole-on model.
Here, the accretion radius is $r_{\infty (i)}$ 
of the outermost streamline satisfying the accretion condition.
The accretion rate is obtained with using the 
accretion radius as $\dot{M} = \pi r_{\rm acc}^2 \rho_\infty v_\infty$.

As shown in figure \ref{fig7},
the accretion radius tends to decrease as $\Gamma'$ increases.
This is because the strong radiation force works to push the gas 
and prevent the accretion.
However, since the radiation force is weakened by attenuation,
accretion is possible even for super-Eddington regime ($\Gamma'>1$)
for the case of $\tau_{\rm HL}=1.0$ and $3.0$.
The effect of attenuation is more effective for larger $\tau_{\rm HL}$,
the accretion radius is larger for $\tau_{\rm HL}=3.0$ 
than for $\tau_{\rm HL}=1.0$.
In the optically thin limit,
the accretion radius of the isotropic model is 
larger than that of the pole-on model,
since the radiation flux is enhanced at around the rotation axis
of the disk ($z$-axis), effectively pushing the gas coming from the $z$ direction.
However, the difference between the isotropic model and the pole-on model
becomes gradually smaller with an increase of $\tau_{\rm HL}$
and $r_{\rm acc}$ for the pole-on model exceeds that for the isotropic model
in the case of $\tau_{\rm HL} \gtrsim 1.0$
(see the \S \ref{Structure of flow} for detailed reason).

It is found that the accretion radius for $\tau_{\rm HL}=3.3$
rapidly decreases to zero at $\Gamma'\sim 3.4$ for the pole-on model
and at $\Gamma'\sim 2.8$ for the isotropic model.
Also, in the case of $\tau_{\rm HL}=1.0$,
the accretion radius becomes suddenly zero at $\Gamma'\sim 1.3$ for both models.
This can be understood as follows.
In the case of large $\Gamma'$,
the streamlines greatly bend near the center
by the strong radiation force,
and many streamlines merge into one.
The mass transport rate of this single streamline becomes large.
Thus the accretion rate varies greatly 
depending on whether this streamline satisfies the accretion condition or not.
For example, we can see that
the streamline nearest to $z$-axis is responsible for about 80\% 
of the mass transport rate.
If this streamline satisfy the accretion condition,
the accretion rate exceeds $0.8\dot{M}_{\rm HL}$.
In the situation that the accretion condition is not satisfied for this streamline, 
other streamlines does not satisfy the condition
since their reach points are more distant.
Thus, the accretion rate becomes zero.

\section{Asymmetry of accretion cross-section}
\label{asymmetry}
\begin{figure}[t]
  \begin{center}
    \includegraphics[width=88mm]{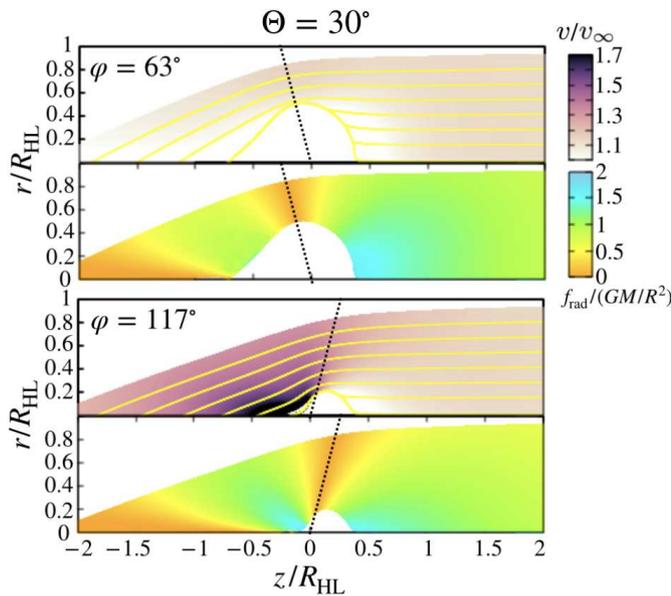}
  \end{center}
  \caption{
  Same as figure \ref{fig2} but for $\Theta=30^{\circ}$, where $\tau_{\rm HL}=0.33$ and $\Gamma'=1.0$ are employed. The upper and lower panels show results on the plane of $\varphi=63^{\circ}$ and $117^{\circ}$. The dotted lines indicate the direction of the disk plane.  

}\label{fig8}
\end{figure}

In figure \ref{fig3},
we find that the accretion cross section for $\Theta=30^\circ$ and $60^\circ$
is not symmetrical to the plane of $\varphi=90^\circ$.
The cross section for $\varphi<90^\circ$ is larger than that for $\varphi>90^\circ$
for the case of $(\tau_{\rm HL},\Gamma')=(0.33, 1.0), (1.0, 1.0)$ and $(1.0, 2.0)$.
Such a trend is understood 
in figure \ref{fig8},
which is the same as figure \ref{fig2}
but for $(\tau_{\rm HL}, \Gamma')=(0.33,1.0)$.
Upper and lower panels show the flow structure 
of $\varphi=63^\circ$ and 
$\varphi=117^\circ$, respectively.
The dotted lines indicate the direction of the disk plane.
The region where the radiation force exceeds gravity
is larger in figure \ref{fig2} than in figure \ref{fig8}.
This is due to the small optical thickness ($\tau_{\rm HL}=0.33$).

In the case of $\varphi=63^\circ$ (upper panel),
the flow is closest to the rotation axis of the disk 
in the region of $z>0$.
In this region, the radiation force 
which is much stronger than the gravity
effectively reduces the flow velocity.
The streamlines near the $z$-axis are bent significantly
and many streamlines merge at $0\lesssim z/R_{\rm HL} \lesssim 0.5$.
At the region of $z<0$,
the radiation force does not exceed the gravity
since this region is relatively close to the disk plane (dotted line), 
and since the flow is very optically thick.
Hence, the gas reaches the $z$-axis 
while being slowed down by the gravity.

The flow is effectively slowed down 
even in the case of $\varphi=117^\circ$ (bottom panel).
However, the flow passes through the disk plane (dotted line)
at the the region of $z>0$.
Thus, the flow tends to be less decelerated 
than in the case of $\varphi=63^\circ$.
Due to the relatively weak radiation force, 
the bending of streamlines is small 
and the number of merging streamlines is small.
The flow approaches the rotation axis of the disk 
in the region of $z<0$.
The strong radiation force, which is stronger than the gravity,
accelerates the flow so that 
the gas reaches the $z$-axis 
at a relatively high speed.

As described above, 
the difference in radiation force leads to the difference 
in the velocity distribution.
The flow velocity of $\varphi=63^\circ$
is lower than that of $\varphi=117^\circ$,
so that 
the accretion condition is more likely 
to be satisfied in the case of $\varphi=63^\circ$.
Thus, the accretion cross section 
becomes wider for $\varphi < 90^\circ$
than for $\varphi > 90^\circ$.

\def\thesection{Reference\Alph{section}}

 \end{document}